\documentclass[
    ,final            % use final for the camera ready runs
  ]
  {aipproc}

\layoutstyle{6x9}

\usepackage[utf8]{inputenc}

\def\nh{$N_{\mathrm H}$}

\begin{document}

\title[Extragalactic Hard X-ray Surveys]{Extragalactic Hard X-ray Surveys: From \emph{INTEGRAL} to \emph{Simbol-X}}


\author{S. Paltani}{
  address={ISDC, University of Geneva, Versoix, Switzerland}
}

\author{T. Dwelly}{
  address={School of Physics and Astronomy, University of Southampton, Southampton UK}
}

\author{R. Walter}{
  address={ISDC, University of Geneva, Versoix, Switzerland}
}

\author{A. P. Marscher}{
  address={Institute for Astrophysical Research, Boston University, Boston, USA}
}

\author{S. G. Jorstad}{
  address={Institute for Astrophysical Research, Boston University, Boston, USA}
}

\author{I. M. McHardy}{
  address={School of Physics and Astronomy, University of Southampton, Southampton UK}
}

\author{T. J.-L. Courvoisier}{
  address={ISDC, University of Geneva, Versoix, Switzerland}
}

\begin{abstract}
 We present some results of the deepest extragalactic survey performed by the INTEGRAL satellite. The fraction of very absorbed AGN is quite large. The sharp decrease in the absorption fraction with X-ray luminosity observed at lower-energy X-rays is not observed. The current lack of truly Compton-thick objects, with an upper limit of 14\% to the size of this population, is just compatible with recent modeling of the cosmic X-ray background.

We also study the prospects for a future hard X-ray serendipitous survey with Simbol-X. We show that \textit{Simbol-X} will easily detect a large number of serendipitous AGN, allowing us to study the evolution of AGN up to redshifts about 2, opening the door to the cosmological study of hard X-ray selected AGN, which is barely possible with existing satellites like Swift and INTEGRAL.
\end{abstract}

\maketitle

%%%%%%%%%%%%%%%%%%%%%%%%%%%%%%%%%%%%%%%%%%%%
%% MAINMATTER
%%%%%%%%%%%%%%%%%%%%%%%%%%%%%%%%%%%%%%%%%%%%

\section{Introduction}
Since the discovery of a tight connection between supermassive black holes and their host galaxies \cite{MagoEtal-1998-DemMas}, AGN surveys have become an important part of observational cosmology. Thanks to the excellent sensitivity of current satellites, the X-ray domain below 10\,keV is very efficient at targeting AGN. However, this energy domain is very sensitive to absorption and it is expected that X-ray surveys may provide only a partial view of the AGN population. A large population of strongly absorbed AGN has indeed long been expected \cite{SettWolt-1989-ActGal} on the basis of the discovery by the HEAO-1 satellite of an apparently diffuse X-ray emission at high galactic latitude \cite{MarsEtal-1980-DifXra}. Recent modeling of this so-called cosmic X-ray background (CXB) emission \cite{GillEtal-2007-SynCos} even requires a significant fraction of objects with hydrogen column densities \nh\ larger than $\sim 10^{24}$\,cm$^{-2}$. These Compton-thick objects emit very little radiation below 10\,keV and thus require deep X-ray observations. To sample efficiently these objects, it is necessary to perform surveys in the hard X-ray domain ($>\sim 20$\,keV), where photoelectric absorption is much smaller. Such surveys are now possible with INTEGRAL and SWIFT and the hard X-ray selected AGN population has been studied over the full sky \cite{SazoEtal-2007-HarXra,TuelEtal-2008-SwiBat}. However, the sensitivity limit in the hard X-rays is still very poor. We present here the absorption properties the sources found in the most sensitive high-latitude hard X-ray survey to date, an INTEGRAL survey covering a $\sim 2500$\,deg$^2$ region around 3C~273 and the Coma cluster \cite{PaltEtal-2008-DeeInt}.  We also study the potential of Simbol-X for cosmological studies of AGN selected in the hard X-rays compared to what will be achieved by INTEGRAL and Swift at the end of their missions.

\section{The INTEGRAL 3C~273/Coma survey}
The quasar 3C~273 has been the target of several INTEGRAL observations. Adding an observation of the nearby Coma cluster, this region is the most exposed extragalactic region, consisting of more than 1600 pointings for a total exposure time of close to 4\,Ms. We used these data to build a 3000x3000-pixel mosaic in the 20--60\,keV energy range. About 2500\,deg$^2$ have been observed for more than 10\,ks effective exposure time, the threshold above which systematics are averaged out. We find 34 candidate sources at the 5$\sigma$ level. In the center of the mosaic, this corresponds to about $5\times 10^{-12}$\,erg cm$^{-2}$ s$^{-1}$. Details of the analysis are presented in \cite{PaltEtal-2008-DeeInt}.

\section{Absorption properties}
17 of the 34 candidate sources had adequate X-ray measurements allowing us to determine their hydrogen column density, \nh. We have started a follow-up program with Swift/XRT which allowed us to determine \nh\ in 7 additional sources. For the 10 remaining objects, 3 have a counterpart in the ROSAT all-sky survey bright source catalog (RASS-BSC) \cite{VogeEtal-1999-ROSBSC}, allowing us to obtain a crude estimate of \nh. For the 7 sources without detected counterpart in the RASS-BSC, \nh\ is roughly constrained within the range $2\times 10^{22}$--$10^{25}$\,cm$^{-2}$.

\begin{figure}
  \includegraphics[height=.21\textheight]{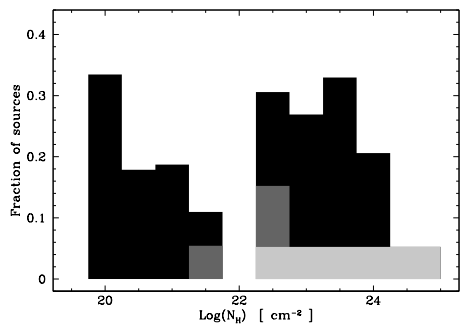}
  \caption{\label{nh_dist}\nh\ distribution in the 3C273/Coma survey. Actual \nh\ measurements are shown in black; medium grey shows \nh\ estimates based on the source flux in the RASS-BSC; light grey corresponds to sources without counterparts in the RASS-BSC and are thus evenly distributed between $2\times 10^{22}$ and $10^{25}$\,cm$^{-2}$.}
\end{figure}
Figure~\ref{nh_dist} shows the distribution of the intrinsic hydrogen column density \nh\ for the 34 sources. The fraction of objects with \nh$>10^{22}$\,cm$^{-2}$ is found to be about 60\%; if one discards the 7 sources without X-ray counterparts, this figure becomes 52\%, making it a stringent lower limit. This is significantly larger than the fraction observed in X-ray surveys below 10\,keV in the local universe \cite{Hasi-2008-AbsPro}, showing the capabilities of hard X-ray surveys to detect strongly absorbed objects. The absence of confirmed Compton-thick objects (\nh$>10^{24}$\,cm$^{-2}$) puts however an upper limit to their fraction of 14\%, consistent with previous hard X-ray surveys \cite{TuelEtal-2008-SwiBat}. A recent modeling of the CXB \cite{GillEtal-2007-SynCos} predicted a fraction of AGN with \nh$>10^{22}$\,cm$^{-2}$ of 65\% at the fluxes sampled by our survey, which is perfectly compatible with our measurement. Their expected fraction of Compton-thick AGN (15\%) is just compatible with our upper limit, and the failure to find any Compton-thick object up to now points to a possible inconsistency.

The decrease in the absorbed fraction with luminosity observed in several X-ray surveys below 10\,keV \cite{Hasi-2008-AbsPro} is not observed here. The significance of this discrepancy is only about 2$\sigma$ and may therefore be due to the small sample size.

\section{Prospects for future hard X-ray surveys}

Over the coming years, Swift, which is primarily dedicated to the observation of gamma-ray bursts, will keep performing a relatively uniform survey of the full high-latitude sky. INTEGRAL, although it spends most of the time looking at the galactic plane, may spend a significant amount of time in a small high-latitude sky region, in order to reach the highest sensitivity possible with the current generation of instruments. Because of its small field of view, Simbol-X is not a survey instrument. However, thanks to its extraordinary sensitivity, it will provide a very interesting serendipitous survey of the fields around extragalactic targets. Here, we compare the expected results of these three different kinds of survey, making rough assumptions on the survey properties.

Swift will eventually reach fluxes twice as low as in \cite{TuelEtal-2008-SwiBat}, i.e.\ about $5\times 10^{-12}$ erg s$^{-1}$ cm$^{-2}$ in the rest frame 20--60 keV band, over about 30\,000 deg$^2$. The deepest coded-mask hard X-ray survey could be obtained by spending some 20\,Ms in a single $5\times 5$ dithering pattern with \textit{INTEGRAL}, reaching a limiting flux of $2\times 10^{-12}$ erg s$^{-1}$ cm$^{-2}$ in the rest frame 20--60 keV band over 1000 deg$^2$. The properties of a serendipitous Simbol-X survey are more difficult to guess. It seems reasonable to assume typical observing times of 100\,ks and that there will be about 200 pointings, making total usable sky area of 5 deg$^2$. The projected limiting flux will be $2\times 10^{-14}$ erg s$^{-1}$ cm$^{-2}$ in the rest frame 20--60 keV band, well above the source confusion limit.

We simulate a cosmological population of AGN in the hard X-rays based on the local 20--60\,keV luminosity function (LF) observed in \cite{PaltEtal-2008-DeeInt}. For simplicity we assume a pure density evolution close to that observed in the 2--10\,keV domain ($\sim (1+z)^4$) \citep{UedaEtal-2003-CosEvo}. Swift will detect roughly 650 sources, while some 60 sources will be found in the INTEGRAL ultra-deep survey. Simbol-X will detect serendipitously about 1200 sources. The Swift redshift distribution peaks below $z=0.03$, while the INTEGRAL is less peaked at about $z=0.05$ wih a tail extending beyond $z=0.1$. On the other hand, pratically all AGN in the Simbol-X survey lie beyond $z=0.1$, with a peak at $z\sim 0.5$ and a tail extending beyond $z=2$.
\begin{figure}
\includegraphics[height=.24\textheight]{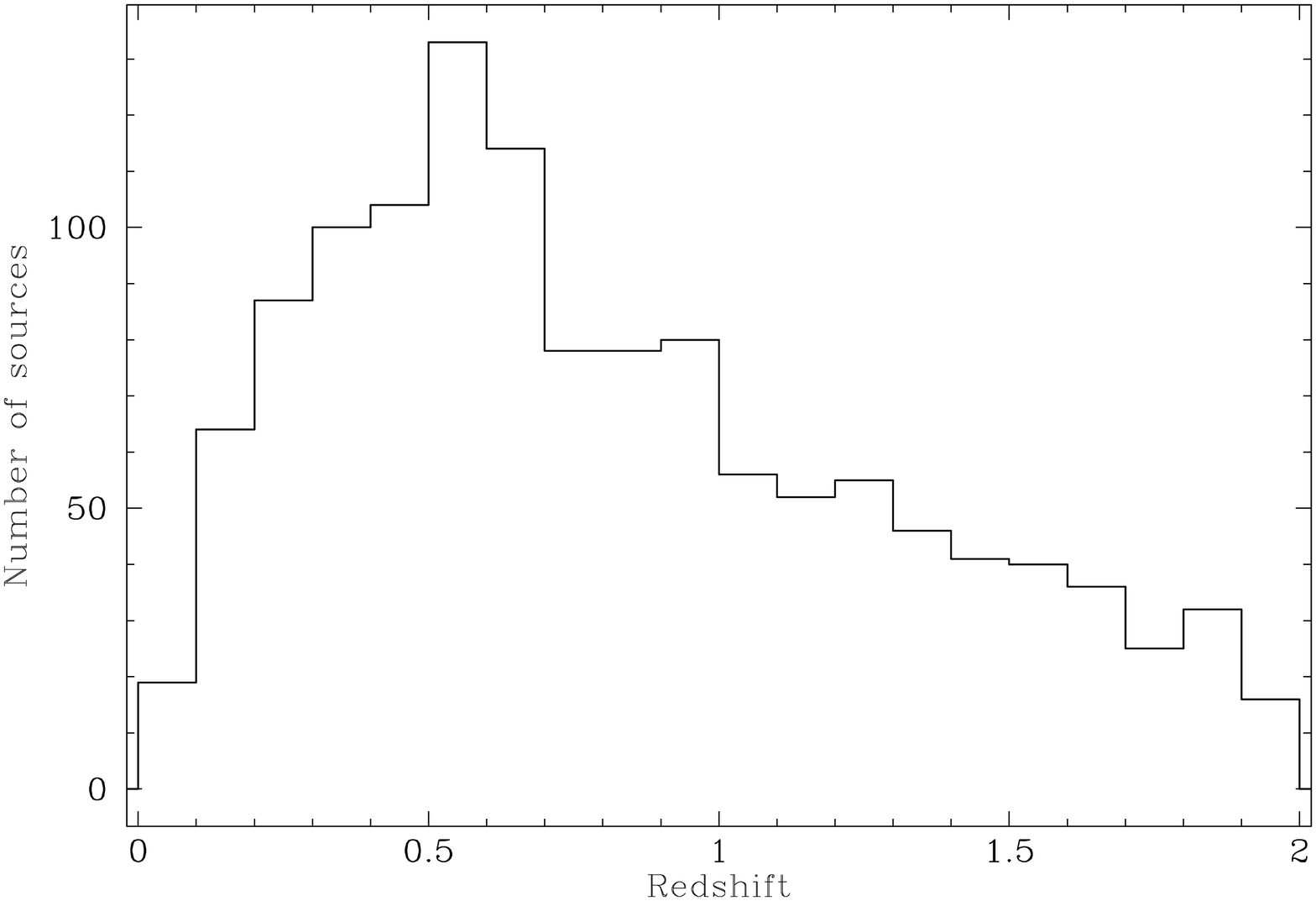}\hspace{5mm}\includegraphics[height=.24\textheight]{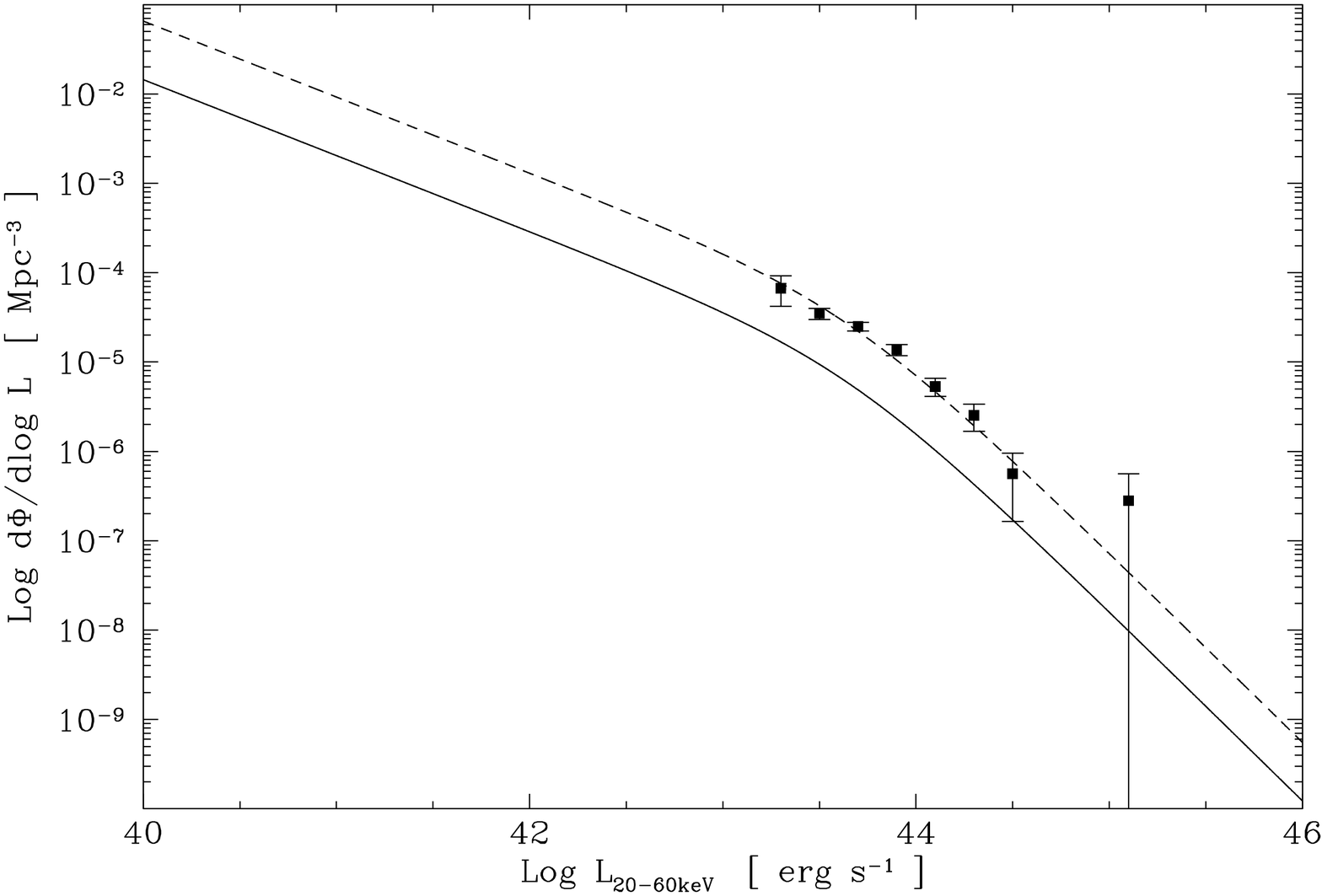}
  \caption{Left: expected redshift distribution for the Simbol-X serendipitous hard X-ray survey. Right: hard X-ray AGN LF from the Simbol-X serendipitous survey in the redshift range $0.4\le z<0.6$. The solid line is the local LF; the dashed line shows the density evolution $\sim (1+z)^4$.}
\end{figure}

While Swift will be able to determine the local hard X-ray LF with good accuracy below $z=0.05$, the INTEGRAL ultra-deep survey is needed to study the low-luminosity tail up to $z=0.1$. The LF from the Simbol-X survey is very good up to $z\sim 1$, probing the high-luminosity tail up to $z\sim 2$. A LF density evolution such as that observed in 2--10\,keV surveys \cite{UedaEtal-2003-CosEvo} will be very easily detected by Simbol-X. This can be used to constrain the model of AGN evolution in the hard X-rays, and to compare with that determined at lower energies.

\section{Conclusions}
The survey we presented here is currently the deepest extragalactic hard X-ray survey. A complete follow-up campaign with Swift/XRT is under way. We find a large fraction of very absorbed object, but we do not find any truly Compton-thick AGN, putting an upper limit to this population any 14\%, which is just compatible with expectations from recent CXB synthesis models \cite{GillEtal-2007-SynCos}.

We study the prospects of future extragalactic hard X-ray surveys: an all-sky Swift survey, an ultra-deep INTEGRAL survey and a serendipitous Simbol-X survey. Making reasonable assumptions about the characteristics of these surveys, we simulate populations of AGN based on the current knowledge of AGN evolution and construct their LFs. We find that Swift will make an excellent determination of the local hard X-ray LF, but a ultra-deep INTEGRAL survey is needed to study the LF beyond $z=0.05$ and up to $z\sim 0.1$. However, Simbol-X will open the door to cosmological studies of hard X-ray selected AGN populations, with an expected redshift distribution peaking around $z\sim 0.5$ and a tail extending beyond $z=2$. At this redshift, Simbol-X will study the evolution of the high-luminosity tail of the LF.



\bibliographystyle{aipprocl} % if natbib is missing

\bibliography{biblio}

\IfFileExists{\jobname.bbl}{}
 {\typeout{}
  \typeout{******************************************}
  \typeout{** Please run "bibtex \jobname" to optain}
  \typeout{** the bibliography and then re-run LaTeX}
  \typeout{** twice to fix the references!}
  \typeout{******************************************}
  \typeout{}
 }

\end{document}